\documentclass[12pt,epsf]{article}
\usepackage[dvips]{epsfig}
\newcommand{\rem}[1]{}

\catcode`\@=11

\title{\bf
A normal form for excitable media}
\author 
{
{\em } Georg A. Gottwald$^1$ and Lorenz Kramer$^2$\footnote{LK
(1941-2005) requiescat in pace.} \\
{\em $^1$ School of Mathematics \& Statistics, University of
Sydney,}\\
{\em NSW 2006, Australia.}\\
{\small gottwald@maths.usyd.edu.au}\\
{\em $^2$ Physikalisches Institut, Universit\"at Bayreuth, }\\
{\em Universit\H atstra{{\ss}}e 30, D-95440, Bayreuth, Germany.}\\
}
\date{}

\begin{document}

\begin{titlepage}
\setcounter{page}{1}

\vfill
\maketitle

\begin{abstract}
\noindent  
We present a normal form for travelling waves in one-dimensional
excitable media in form of a differential delay equation. The normal
form is built around the well-known saddle-node bifurcation
generically present in excitable media. Finite wavelength effects are
captured by a delay. The normal form describes the behaviour of
single pulses in a periodic domain and also the richer behaviour of
wave trains. The normal form exhibits a symmetry preserving Hopf
bifurcation which may coalesce with the saddle-node in a
Bogdanov-Takens point, and a symmetry breaking spatially inhomogeneous
pitchfork bifurcation. We verify the existence of these bifurcations
in numerical simulations. The parameters of the normal form are
determined and its predictions are tested against numerical
simulations of partial differential equation models of excitable media
with good agreement.
\end{abstract}
\vfill

\end{titlepage}


\noindent
{\bf{ Excitable media are often found in biological and chemical
systems. Examples of excitable media include electrical waves in
cardiac and nerval tissue \cite{WinfreeBook,Davidenko}, cAMP waves in
slime mold aggregation \cite{dictyostelium} and intracellular calcium
waves \cite{calcium}. Excitable media support localized pulses and
periodic wave trains. In 2 dimensions rotating vortices (or spirals)
and in 3 dimensions scroll waves
\cite{Winfree,Winfree90,Winfree94,Margerit01,Margerit02} are possible. The
critical behaviour of pulses, wave trains and spirals,
i.e. propagation failure, is often associated with clinical
situations. The study of spiral waves is particularly important as
they are believed to be responsible for pathological cardiac
arrhythmias \cite{Chaos}. Spiral waves may be created in the heart
through inhomogeneities in the cardiac tissue. Some aspects of spiral
wave break up can be studied by looking at a one-dimensional slice of
a spiral i.e. at a one-dimensional wave train
\cite{Courtemanche}.\\ \noindent We investigate critical behaviour
relating to one-dimensional wave propagation. We develop a normal form
which allows us to study the bifurcation behaviour of critical
waves. In particular, the normal form predicts a Hopf bifurcation and
a symmetry breaking pitchfork bifurcation. The symmetry breaking
pitchfork bifurcation can be numerically observed as an instability
where every second pulse of a wave train dies. This seems to be
related to alternans \cite{Nolasco,Karma_A}, which are discussed in
the context of cardiac electric pulse propagation. }}


\section{Introduction}
\label{Sec-intro}
Many chemical and biological systems exhibit excitability. In small
(zero-dimensional) geometry they show threshold behaviour, i.e. small
perturbations immediately decay, whereas sufficiently large
perturbations decay only after a large excursion. This behaviour is
crucial for the electrical activation of cardiac tissue or the
propagation of nerve pulses where activation should only be possible
after a sufficiently large stimulus. Moreover, the decay to the rest
state allows for the medium to be repeatedly activated - also crucial
for the physiological functioning of the heart and the nervous
system. One-dimensional excitable media support travelling pulses, or
rather, periodic wave trains ranging in wavelength $L$ from the
localized limit $L \to \infty$ to a minimal value $L_c$ below which
propagation fails. Pulses and wave trains are best-known from nerve
propagation along axons. In two dimensions one typically observes
spiral waves. Spirals have been observed for example in the
auto-catalytic Belousov-Zhabotinsky reaction
\cite{Winfree}, in the aggregation of the slime mold dictyostelium
discoideum \cite{dictyostelium} and in cardiac tissue
\cite{Davidenko}.

For certain system parameters the propagation of isolated pulses and
wave trains may fail (see for example \cite{synergetics,Jahnke}). The
analytical tools employed to describe these phenomena range from
kinematic theory
\cite{Zykov,Mikhailov1}, asymptotic perturbation theory
\cite{Meron,Karma1,Karma2} to dynamical systems approaches
\cite{Barkley94,Melbourne,GottwaldKramer04}. Numerical observations
reveal that independent of the detailed structure of a particular
excitable media model, the bifurcation behaviour of excitable media is
generic. For example, a saddle-node bifurcation is generic for single
pulses and for wave trains. However, there exists no general theory
which accounts for all bifurcations which may appear. In this paper we
develop a normal form for excitable media which is build around the
observation that the propagation failure of a one-dimensional wave
train is mediated by the interaction of a pulse with the inhibitor of
the preceding pulse.\\

\noindent
In Section~\ref{Sec-1D} we briefly review some basic properties of
excitable media and illustrate them with a specific example. In
Section~\ref{Sec-NF} we introduce the normal form. The properties and
the bifurcation scenarios of this normal form are investigated in
Section~\ref{Sec-BT}. In Section~\ref{Sec-Num1} we show how the
parameters of the normal form can be determined from numerical
simulations of the excitable medium and then compare the predictions
of the normal form with actual numerical simulations of a partial
differential equation. The paper concludes with a discussion in
Section~\ref{Sec-Disc}.


\section{One-dimensional excitable media}
\label{Sec-1D}
Most theoretical investigations of excitable media are based on coupled
reaction-diffusion models. We follow this tradition and investigate a
two-component excitable medium with an activator $u$
and a non-diffusive inhibitor $v$ described by
\begin{eqnarray}
\label{barkley}
\partial_t u &=& D u_{xx} +{\cal{F}}(u,v),\quad {\cal{F}}(u,v) = u(1-u)(u-u_s-v) \nonumber \\
\partial_t v &=&  \epsilon \ (u- a\  v)\  .
\end{eqnarray}
This is a reparametrized version of a model introduced by Barkley
\cite{Barkley91}. Note that the diffusion constant $D$ is not a
relevant parameter as it can be scaled out by rescaling
length. Although the normal form which we will introduce in
Section~\ref{Sec-NF} is independent of the particular model used, we
illustrate some basic properties of excitable media using the
particular model (\ref{barkley}). Later in Section~\ref{Sec-Num1} we
show correspondence of the predictions of our normal form with
numerical simulations of the model (\ref{barkley}). Our choice of
model is motivated by the fact that this model incorporates the
ingredients of an excitable system in a compact and lucid way. Thus,
for $u_s>0$ the rest state $u_0=v_0=0$ is linearly stable with decay
rates $\sigma_1=u_s$ along the activator direction and
$\sigma_2=\epsilon a$ along the inhibitor direction. Perturbing $u$
above the threshold $u_s$ (in 0D) will lead to growth of $u$.  In the
absence of $v$ the activator would saturate at $u=1$ leading to a
bistable system.  A positive inhibitor growth factor $\epsilon$ and
$a>0$ forces the activator to decay back to $u=0$. Finally also the
inhibitor with the refractory time constant $(\epsilon\ a)^{-1}$ will
decay back to $v=0$. For $a>1/(1-u_s)$ the system is in
zero-dimensional systems no longer excitable but instead bistable.

In order to study pulse propagation in one-dimensional excitable media
it is useful to first consider the case of constant $v$. The resulting
bistable model is exactly solvable \cite{Keener} and the pulse
velocity is $c_f(v)=\sqrt{\frac{D}{2}}[1-2(u_s+v)]$. Hence,
excitability requires that $u_s$ is below the stall value
$\frac{1}{2}$. The quantity $\Delta=\frac{1}{2}-u_s$ characterizes the
strength of excitability and $c_f(0)$ coincides with the solitary
pulse velocity for $\epsilon
\to 0$.

Clearly, for $u_s<u_c=\frac{1}{2}$ and not too large $a$, pulse
propagation fails for $\epsilon$ larger than some $\epsilon_c$. The
critical growth factor $\epsilon_c$ marks the onset of a saddle-node
bifurcation
\cite{Zykov,Meron,GottwaldKramer04}. The saddle-node can be intuitively
understood when we consider the activator pulse as a heat source, not
unlike a fire-front in a bushfire. Due to the inhibitor the width of
the pulse decreases with increasing $\epsilon$. Hence, the heat
contained within the pulse decreases. At a critical width, or critical
$\epsilon$, the heat contained within the pulse is too small to
ignite/excite the medium in front of the pulse.

\begin{figure}
\label{Fig-pulses}
\centerline{
\psfig{file=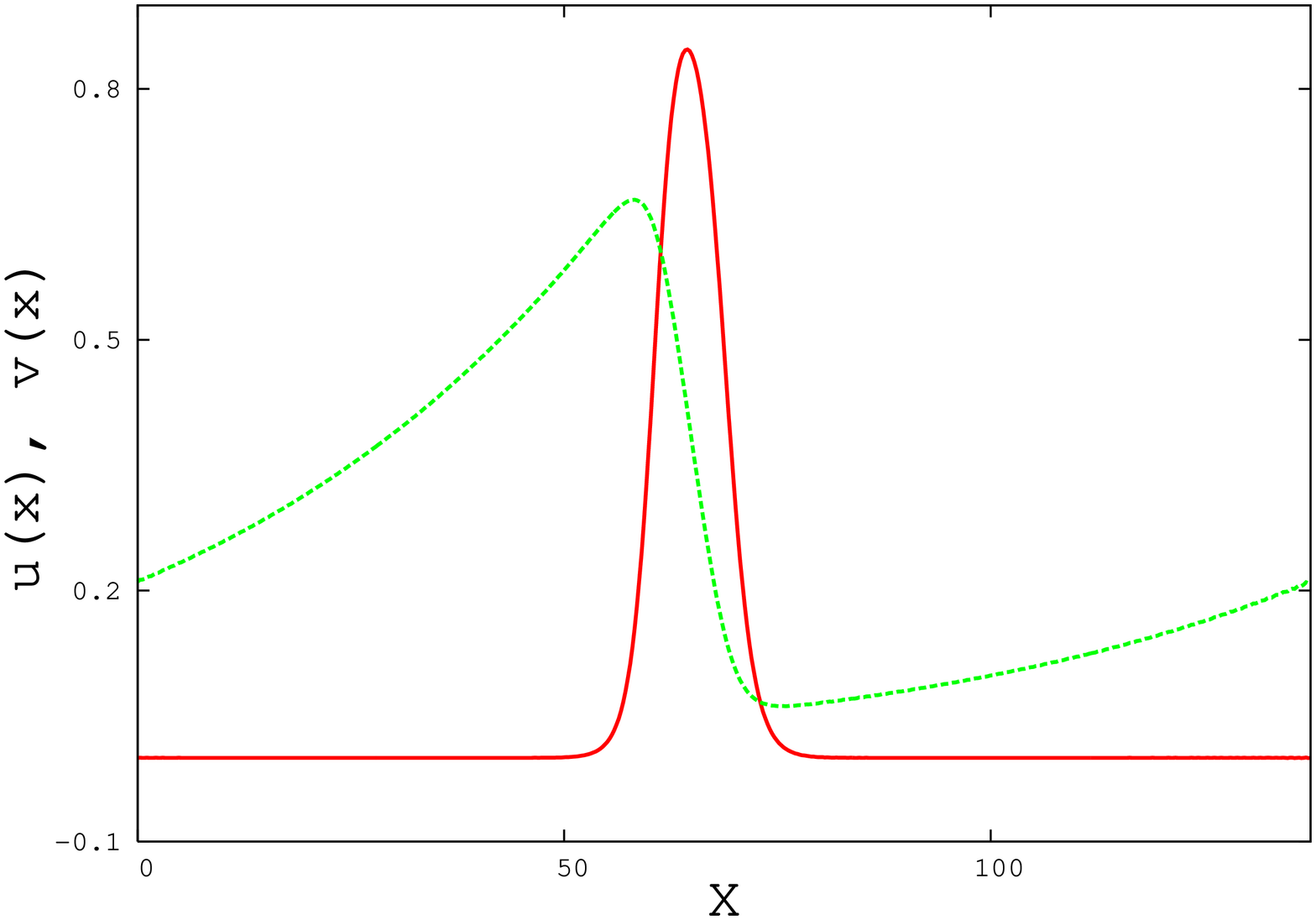,angle=270,width=3.0in}}

\centerline{(a)}

\centerline{
\psfig{file=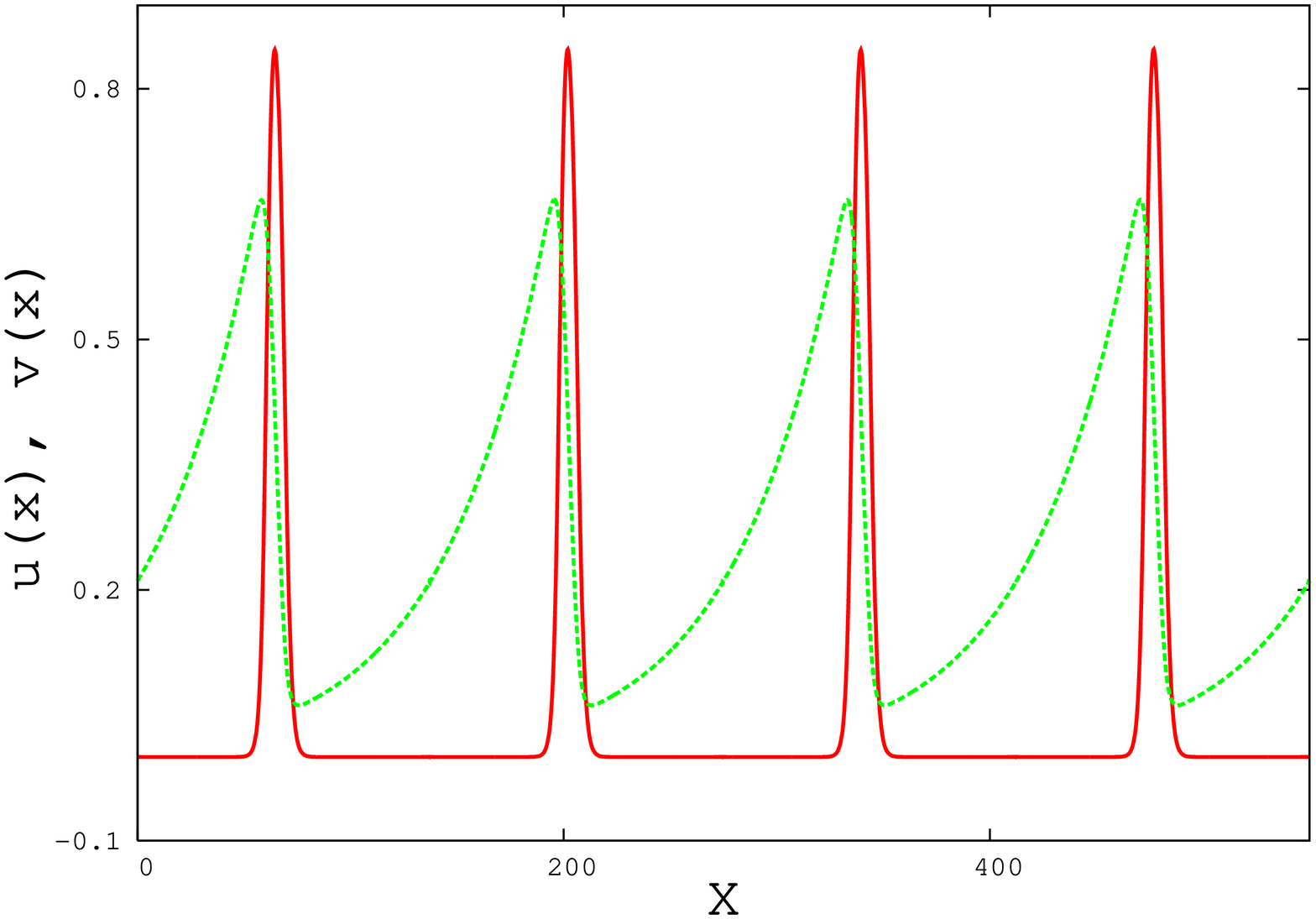,angle=270,width=3.0in}}

\centerline{(b)}
\caption{Typical profiles of the activator $u$ (continuous
line) and inhibitor $v$ of close-to-critical travelling wave solutions
in a one-dimensional ring of (\ref{barkley}). Parameters are
$a=0.22,u_s=0.1,\epsilon=0.035$. (a) Single pulse with $L=137.5$. (b)
Wave train with $L=550$.}

\end{figure}

Even if a given set of equation parameters allows for propagation of a
single isolated pulse, the system may not necessarily support a pulse
in a periodic box of finite length or a wave train consisting of
several of such pulses: If the distance $L$ between two consecutive
pulses of the train becomes too small, the pulses run into the
refractory tail of the preceding pulse (see Figure~\ref{Fig-pulses}),
and may consecutively decay . Hence, propagation failure for periodic
wave trains is controlled by the decay of the inhibitor, and
propagation is only possible when the inter-pulse distance $L$ is
larger than a critical wavelength $L_c$. Note that $L_c$ diverges for
$a \to 0$ when the decay rate of the inhibitor $\sigma_2$
vanishes. The critical wavelength $L_c$ is a lower bound for the
wavelength for the existence of periodic wave trains. One can also
think of keeping $L$ fixed and, as before, vary $\epsilon$. Then the
saddle-node $\epsilon_c(L)$ is a monotonically increasing function.

\noindent
In the next Section we present a normal form which incorporates the
saddle-node bifurcation, and moreover predicts other types of
bifurcations.


\noindent
\section {The normal form}
\label{Sec-NF}
\vskip 5pt 

It is well known that an isolated pulse undergoes a saddle-node
bifurcation above a certain threshold value of the refractoriness
$\epsilon_c$. The corresponding generic normal form for such a
saddle-node bifurcation is given by
\begin{eqnarray}
\label{SN0}
\partial_t X = -\mu - g X^2\; ,
\end{eqnarray}
where $X$ is, for example, the amplitude or velocity of a pulse with
the corresponding values at the saddle-node bifurcation
subtracted. The bifurcation parameter $\mu$ measures the distance from
the bifurcation point and is proportional to
$\epsilon-\epsilon_c$. The normal form (\ref{SN0}) accurately
describes the behaviour of isolated solitary pulses in one-dimensional
excitable media close to criticality. We use this normal form for a
saddle-node bifurcation for an isolated pulse (\ref{SN0}) as a seed to
construct a normal form for general travelling in excitable media
incorporating finite wavelength effects.

In particular we look at a single pulse on a ring with finite length,
i.e. in a one-dimensional box with periodic boundary conditions, and
at wave trains with a finite wavelength. Both cases are depicted in
Figure~\ref{Fig-pulses}. In that case the saddle-node (\ref{SN0}) will
be disturbed and will depend on the length of the periodic box in the
case of a single pulse (Figure~\ref{Fig-pulses}a) or on the wavelength
of the wave train (Figure~\ref{Fig-pulses}b). The interaction of the
pulse (or a member of a wave train) with the preceding pulse (or more
accurate with its inhibitor; see Figure~\ref{Fig-pulses}) modifies as
discussed above the bifurcation behaviour. We may therefore extend the
saddle-node normal form to
\begin{eqnarray*}
\label{SN1}
\partial_t X = -\mu - g X^2 - \beta_0 V(t-\tau)\; ,
\end{eqnarray*}
where $V(t-\tau)$ describes the inhibitor of the preceding pulse who
is temporally displaced by $\tau=L/c_0$ where $L$ is the wavelength of
the pulse train, i.e. the distance between two consecutive pulses, and
$c_0$ is its uniform velocity. We neglect here a possible temporal
dependency of $\tau$. Note that in the case of a single pulse in a
ring, $V(t-\tau)$ describes the inhibitor of the single pulse which
had been created by the pulse at the time of the last revolution
around the ring and $L$ is simply the length of the periodic box.\\
\noindent
We assume an exponential decay (in space and time) of the inhibitor of
well separated pulses. This is the case for the system
(\ref{barkley}). We may write $V(t-\tau)=\exp(-k
\tau)h(X(t-\tau))$ where the function $h$ and the decay-rate $k$
depend on the particular model chosen; for example for the model
(\ref{barkley}) we have $k=\epsilon a$. In the limiting case of
isolated pulses we note that $\tau \rightarrow
\infty$ and $V(t-\tau)\rightarrow 0$, and hence we retrieve the
unperturbed saddle-node bifurcation (\ref{SN0}). The ansatz for
$V(t-\tau)$ is a simplification where we ignore the cumulative effect
of the inhibitor. (Note that the equation for the inhibitor $v$ in
(\ref{barkley}) can be solved directly and involves an integral over
$u$, i.e. involves 'history'.) The unknown function $h(X(t-\tau))$ can
be Taylor-expanded around the saddle-node $X=0$.

\noindent
We summarize and arrive at the following normal form
\begin{eqnarray*}
\partial_t X = -\mu - g X^2 - \beta (\gamma + X(t-\tau))\; ,
\end{eqnarray*}
where $\beta=\beta_0\exp(-k \tau)$ with $k=\epsilon a$ for the model
equation (\ref{barkley}). Pulses have a nonzero width $\nu$ which
implies that the temporal delay $\tau=L/c_0$ has to be modified to
$\tau=(L-\nu)/c_0$. This equation already produces qualitatively all
the results we will present in the subsequent Sections. However, much
better quantitative agreement is achieved by taking into account that
a variation in the amplitude implies a change in velocities and
henceforth a change of the effective inhibition. If we allow for a
single pulse to have a temporarily varying pulse amplitude or, in the
case of a wave train consisting of distinct members, if we allow for
different amplitudes of individual members of the wave train, we have
to take into account that the propagation behaviour is amplitude
dependent: Larger pulses have larger velocities. Hence, a pulse $X(t)$
which is larger than its predecessor $X(t-\tau)$ runs further into the
inhibitor-populated space created by its predecessor. If
$X(t-\tau)<X(t)$ the finite wavelength induced shift of the
bifurcation is stronger compared to the case of equal
amplitudes. Conversely, if $X(t-\tau)>X(t)$ the finite wavelength
induced shift of the bifurcation is weaker compared to the case of
equal amplitudes. This effect is stronger the larger the difference of
the two amplitudes $X(t-\tau)-X(t)$. The inclusion of the amplitude
differences affects the bifurcation behaviour depending continuously
on the difference $X(t-\tau)-X(t)$. We thus add a term
$\gamma_1(X(t)-X(t-\tau))$ with $0<\gamma_1\ll 1$ into the wavelength
dependant inhibitor term in (\ref{NF}), and arrive at
\begin{eqnarray*}
\partial_t X = -\mu - g X^2 - \beta (\gamma + (1-\gamma_1) X(t-\tau) + \gamma_1X(t))\; ,
\end{eqnarray*}
or after relabeling of $\beta,\gamma,\gamma_1$
\begin{eqnarray}
\label{NF}
\partial_t X = -\mu - g X^2 - \beta (\gamma + X(t-\tau) + \gamma_1X(t))\; .
\end{eqnarray}
It is this equation which we propose as a normal form to study
bifurcations of one-dimensional wave trains. 


\noindent
\section {Properties of the normal form}
\label{Sec-BT}
Before we show how to determine the parameters of the normal form, we
will describe its properties with a main emphasis on
bifurcations. Besides the well-known saddle-node bifurcation we
identify a symmetry preserving Hopf bifurcation and a symmetry
breaking spatially inhomogeneous pitchfork bifurcation. Numerical
integration of partial differential equation models of excitable media
such as (\ref{barkley}) confirm these bifurcation scenarios of the
normal form (\ref{NF}). Although some of these bifurcations have been
previously observed in numerical simulations, up to now there did not
exist a unified framework to study these bifurcations. The normal form
is able to identify these bifurcations as being generic for excitable
media, rather than as being particular to certain models of excitable
media. This is the main achievement of our present work.

\subsection {Saddle-node bifurcation}
\label{Sec-SN}
Numerical simulations of excitable media show that the bifurcations of
a {\it single} propagating pulse in a ring (as in
Figure~\ref{Fig-pulses}a) are different from the bifurcations of a
wave train consisting of several distinct pulses (as in
Figure~\ref{Fig-pulses}b). We first look at a single propagating pulse
before in Section~\ref{Sec-PF} we look at the interaction of different
pulses in a wave train.

\noindent
Equation (\ref{NF}) has the following stationary solutions
\begin{eqnarray}
\label{statio}
{\bar X}_{1,2} =\frac{1}{2g}[-\beta(1+\gamma_1)\pm
\sqrt{\beta^2(1+\gamma_1)^2-4g(\mu+\beta \gamma)}]   \;.
\end{eqnarray}
The upper solution branch is stable whereas the lower one is
unstable. The two solutions coalesce in a saddle-node bifurcation with
\begin{eqnarray}
\label{SN_wt}
{\bar X}_{SN} = -\frac{\beta}{2g}(1+\gamma_1) \quad {\rm{at}} \quad
{\bar \mu}_{SN}= \frac{\beta^2}{4g}(1+\gamma_1)^2-\beta \gamma \;.
\end{eqnarray}
Since $\beta=\beta_0\exp(-\epsilon a \tau)$ is small we have
$\mu_{SN}< 0$. This indicates that the saddle-node of a periodic wave
train occurs at smaller values of the bifurcation parameter $\mu$ than
for the isolated pulse, and the bifurcation is shifted to the left
with respect to the isolated pulse (see
Figure~\ref{Fig-Hopfsketch}). This is a well known fact which we
numerically verified. Note that the limiting case of an isolated
pulses with $L\to
\infty$ implies $\tau=0$ and hence, $\beta=0$. The saddle-node of the
isolated pulse with $X_{SN}=0$ at $\mu=0$ described by (\ref{SN0}) is
recovered.

\begin{figure}
\label{Fig-HopfPDE}
\centerline{
\psfig{file=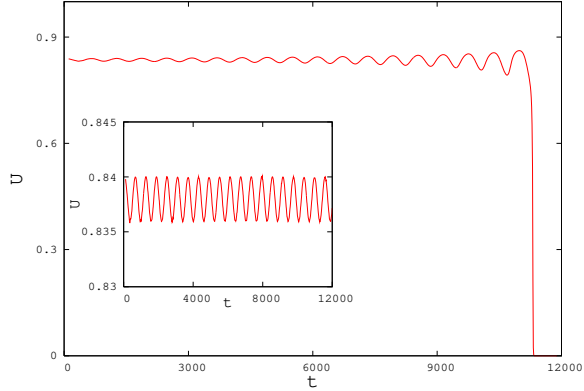,angle=270,width=3.0in}}
\caption{Temporal behaviour of the maximal amplitude $U$ of
the activator $u$ for model (\ref{barkley}) just above the subcritical
Hopf bifurcation. The parameters are $\epsilon=0.3755$ and
$L=246$. The other parameters are as in Figure~\ref{Fig-pulses}. The
subcritical character of the Hopf bifurcation is clearly seen at this
length. If the length $L$ is chosen to be slightly larger and closer
to the actual bifurcation point, the oscillations appear to be stable
for some time as shown in the inset for $L=246.9695>L_c$. However, the
maximal amplitude $U$ for $L=246.9695$ will eventually become visibly
unstable and the pulse will die resulting in $U=0$.}
\end{figure}

Besides this stationary instability the normal form (\ref{NF}) also
allows for a non-stationary bifurcation which we investigate in the
next Section.

\subsection{Symmetry preserving Hopf bifurcation}
\label{Sec-Hopf}
\begin{figure}
\label{Fig-Hopfsketch}
\centerline{
\psfig{file=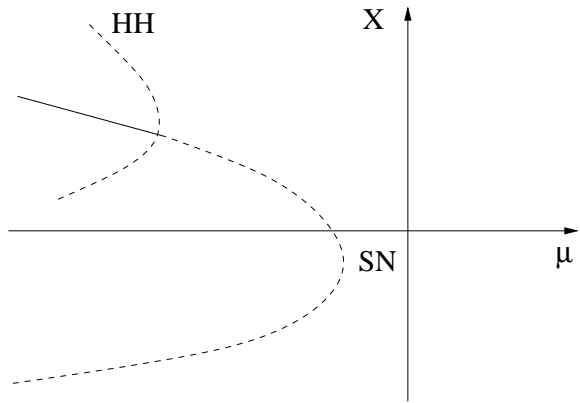,angle=0,width=3.0in}}
\caption{Sketch of the bifurcation diagram for a single
pulse in a ring showing a stationary saddle-node bifurcation (SN) and
a subcritical symmetry preserving Hopf bifurcation (HH).}
\end{figure}

The stability of the homogeneous solution ${\bar X}$ with respect to
small perturbations of the form $\delta X \exp{\sigma t}$ can be
studied by linearizing the normal form around ${\bar X}$. We obtain
\begin{eqnarray}
\label{lin}
\sigma+2g{\bar X} + \beta \gamma_1 + \beta e^{-\sigma \tau} = 0\; .
\end{eqnarray}
Besides the stationary saddle-node bifurcation (\ref{SN_wt}) at
$\sigma=0$ (cf. (\ref{SN_wt})), a Hopf bifurcation $\sigma=i\omega$ is
possible with
\begin{eqnarray}
\label{HomHopfa}
\omega&=&\beta\sin{\omega \tau}\\
\label{HomHopfb}
{\bar X}_{HH}&=& -\frac{\beta}{2g}(\cos{\omega \tau} + \gamma_1)\; .
\end{eqnarray}
In anticipation of the study of wave trains consisting of several
distinct pulses we call this Hopf bifurcation of a single pulse in a
ring a {\it symmetry preserving Hopf bifurcation}. From
(\ref{HomHopfa}) we infer that a Hopf bifurcation is only possible
provided $\beta
\tau>1$, i.e. if the coupling is strong enough and the pulse feels the
presence of the inhibitor of the preceding pulse sufficiently
strong. Since ${\bar X}_{HH} \ge {\bar X}_{SN}$ the symmetry
preserving Hopf bifurcation sets in before the saddle-node
bifurcation, independent of the value of $\beta$. Moreover, the Hopf
bifurcation branches off the upper stable branch of the homogeneous
stationary solutions (\ref{statio}). In Figure~\ref{Fig-Hopfsketch} we
show a schematic bifurcation diagram with the saddle-node bifurcation
and the subcritical Hopf bifurcation for a single pulse in a ring.\\

\noindent
In numerical simulations of the Barkley model and also the
Fitzhugh-Nagumo equations \cite{FHN} we could verify this scenario for
a single pulse in a ring. A Hopf bifurcation had been previously
observed numerically \cite{Knees92} for the Barkley model
\cite{Barkley91} and in \cite{GottwaldKramer04} for the modified Barkley model
(\ref{barkley}). Hopf bifurcations have also been reported to occur in
several other models of excitable media. In
\cite{QuanRudy,Courtemanche,Karma94,Courtemanche96} a Hopf bifurcation
was found in the $8$-variable Beeler-Reuter model
\cite{BeelerReuter}, and in \cite{Karma93,Karma94} in the $4$-variable Noble
model \cite{Noble} and in the $2$-variable Karma model
\cite{Karma93}. We show here that Hopf
bifurcations are generic for travelling waves in excitable media.\\

\noindent
In numerical simulations of model (\ref{barkley}) the Hopf bifurcation
was found to be subcritical. Typical temporal behaviour of the maximal
amplitude of the activator for model (\ref{barkley}) close to the
bifurcation is shown in Figure~\ref{Fig-HopfPDE}. The inset shows the
maximal amplitude slightly above the bifurcation point for about $20$
periods. We counted more than $500$ periods before stability was
visibly lost and the maximal amplitude collapses to zero. We note that
this may have easily lead to the wrong conclusion that the bifurcation
is in fact supercritical rather than subcritical.

\subsection{Bogdanov-Takens point}
The saddle-node and the symmetry preserving Hopf bifurcation coalesce
in a co-dimension-$2$ Bogdanov-Takens bifurcation for $\omega \tau\to
0$. At the Bogdanov-Takens point we have $\beta \tau=1$. The Hopf
bifurcation and the saddle-node bifurcation have been suggested before
to be an unfolding of a Bogdanov-Takens point in \cite{Knees92} and
later in \cite{GottwaldKramer04}. The normal form provides a framework
to study this unfolding. We were able to numerically verify the
condition $\beta
\tau=1$ derived from (\ref{HomHopfa}) by simulating the full partial
differential equation (\ref{barkley}). The parameters $\beta$ and
$\tau$ will be determined further down in Sections~\ref{Sec-Num1}.
We have also numerically simulated the Fitzhugh-Nagumo \cite{FHN}
equations to check that this bifurcation is not particular to our
chosen model (\ref{barkley}).\\

\noindent
The Bogdanov-Takens point is apparent in our normal form (\ref{NF})
and can be derived from it. Close to the saddle-node and the Hopf
bifurcation when $\omega \tau \rightarrow 0$, the dynamics exhibits
critical slowing down. We may therefore expand $X(t-\tau)=X(t)-\tau
\partial_t X(t) + (\tau^2/2) \partial_{tt}X(t) +
{\cal{O}}(\tau^3)$. The normal form (\ref{NF}) becomes at the
Bogdanov-Takens point
\begin{eqnarray}
\label{BT}
\partial_t {\cal{X}} &=& Y\nonumber \\
\partial_t Y &=& -a Y -b {\cal{X}} -\frac{2g}{\tau^2 \beta} {\cal{X}}^2 \; ,
\end{eqnarray}
where ${\cal{X}}=X-{\bar{X}}_{1}$ and ${\bar{X}}_{1}$ satisfies the
stationary version of the normal form (\ref{NF}) and is given by
(\ref{statio}). The linear part of (\ref{BT}) exhibits the correct
eigenvalue structure of a Bogdanov-Takens point. The bifurcation
parameters, $a=2(1-\beta \tau)/(\tau^2\beta)$ and $b =
2(\beta(1+\gamma_1) + 2g{\bar{X}}_{1})/(\tau^2\beta)$, measure the
distance from the Hopf bifurcation and the distance from the
saddle-node bifurcation, respectively.

\subsection{Spatially inhomogeneous pitchfork bifurcation}
\label{Sec-PF}
\begin{figure}
\label{Fig-pdepitch}
\centerline{
\psfig{file=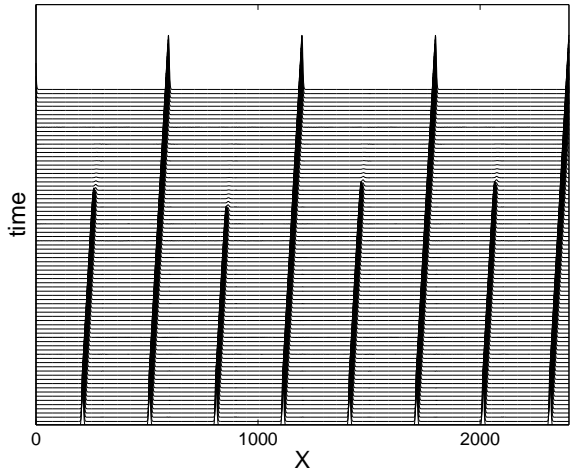,width=3.0in}}
\caption{Space-time plot of the activator $u(x,t)$
demonstrating the spatially inhomogeneous pitchfork
bifurcation. Equation parameters are $\epsilon=0.04897125$ and
$L=300$; all other parameters are as in
Figure~\ref{Fig-pulses}. Initially there is an initial group of $8$
pulses which after some time exhibits a spatial period-doubling
instability and subsequently evolves into a stable propagating wave
train consisting of $4$ remaining pulses.}
\end{figure}

Numerical simulations of systems such as (\ref{barkley}) reveal that a
group of several pulses in a ring do not undergo a symmetry preserving
Hopf bifurcation on increasing the refractoriness $\epsilon$, but
instead develop a symmetry breaking, spatially inhomogeneous
instability whereby every second pulse dies. In
Figure~\ref{Fig-pdepitch} we show an example of such an inhomogeneous
instability. Spatially inhomogeneous bifurcations have been observed
before for periodically paced excitable media
\cite{Hastings00,Echebarria01,Fox02,Henry05}. Here we show that this bifurcation is
generic for wave trains in excitable media and does not require
external pacing.\\

\noindent
The above mentioned inhomogeneous alternating instability is contained
in our normal form. To investigate spatial instabilities we need to
distinguish between consecutive pulses. We may rewrite the normal form
as
\begin{eqnarray}
\label{NFspatial}
\partial_t X_l 
= -\mu - g X_l^2 
- \beta (\gamma + X_{l-1}(t-\tau) + \gamma_1X_l(t))\; ,
\end{eqnarray}
where the subscript $l$ numbers the pulses in a wave train which
interact with their nearest neighbours. Linearizing around the
homogeneous solution $X_l={\bar X}$ according to
\begin{eqnarray*}
X_l = {\bar X} + \delta e^{\sigma t} e^{ipl} \quad {\rm and} \quad X_{l-1} = {\bar X} + \delta
e^{\sigma t} e^{ip(l-1)} 
\end{eqnarray*}
yields as a condition for stationary instabilities (ie. $\sigma=0$)
\begin{eqnarray*}
{\bar X} = -\frac{\beta}{2g}(\gamma_1 + \cos{p} - i \sin{p})\;.
\end{eqnarray*}
Hence stationary instabilities are possible for $p=0$ and for
$p=\pi$. In the homogeneous case $p=0$, the instability is yet again
the spatially homogeneous saddle-node (\ref{SN_wt}). For $p=\pi$ this
is a new type of instability, and we have ${\bar X} = \beta
(1-\gamma_1)/2g$ at the bifurcation point. This spatially
inhomogeneous bifurcation will be identified further down as a
subcritical pitchfork bifurcation. The criterion $p=\pi$ for the
inhomogeneous bifurcation is corroborated by numerical simulations
where every {\it second} pulse dies within a wave train (see
Figure~\ref{Fig-pdepitch}). Note that in traveling wave coordinates of
a partial differential equation model for excitable media, this
instability would correspond to a subcritical period-doubling
bifurcation. \\

\noindent
In order to study this $p=\pi$-bifurcation within our normal form we
need to consider two populations of pulses ($l=1,2$), $X$ and $Y$,
which interact via their inhibitors with each other. We extend our
normal form for the case $p=\pi$ to
\begin{eqnarray}
\label{NF2}
\partial_t X &=&-\mu - g X^2 - \beta (\gamma + Y(t-\tau) +
\gamma_1X(t)) \nonumber \\
\partial_t Y &=& -\mu - g Y^2 - \beta (\gamma + X(t-\tau) + \gamma_1Y(t))\; .
\end{eqnarray}
The system (\ref{NF2}) for wave trains supports two types of
stationary solutions; firstly the homogeneous solution (\ref{statio}),
${\bar X}_h={\bar Y}_h$, which may undergo a saddle-node bifurcation
described by (\ref{SN_wt}). There exists another stationary solution,
an alternating mode, with
\begin{eqnarray}
\label{AI_2}
{\bar X}_{a}=-{\bar Y}_a+\frac{\beta}{g}(1-\gamma_1)\; .
\end{eqnarray}
Associated with this solution is a pitchfork bifurcation at
\begin{eqnarray}
\label{AI_PFa}
\mu_{PF}=\frac{1}{4}\frac{\beta^2(1+\gamma_1)^2}{g}-\frac{\beta^2}{g}-\beta
\gamma 
= {\bar{\mu}}_{SN}-\frac{\beta^2}{g} \le \mu_{SN}\; ,
\end{eqnarray}
when
\begin{eqnarray}
\label{AI_PFb}
X_{PF}=Y_{PF}=\frac{\beta}{2g}(1-\gamma_1)\; .
\end{eqnarray}
Comparing (\ref{SN_wt}) with (\ref{AI_PFa}) shows that the pitchfork
bifurcation sets in before the saddle-node bifurcation. The upper
branch of the homogeneous solution ${\bar X}_h$ given by
(\ref{statio}) at the pitchfork bifurcation point $\mu_{PF}$ coincides
with (\ref{AI_PFb}). Hence the pitchfork bifurcation branches off the
upper branch of the homogeneous solution. It is readily seen that the
pitchfork bifurcation is subcritical because there are no solutions
${\bar X}_{a}$ possible for $\mu > \mu_{PF}$.

We now look at the stability of the homogeneous solution ${\bar X}={\bar Y}={\bar
X}_h={\bar Y}_h$. We study
perturbations $X={\bar X}_h + x\exp{\sigma t}$ and $Y={\bar X}_h +
y\exp{\sigma t}$. Linearization yields as a condition for nontrivial
solutions $x$ and $y$.
%
%
%
\begin{eqnarray}
\label{lin_ai2}
(\sigma+2g {\bar X}_h+\beta \gamma_1) = \pm \beta e^{-\sigma
\tau}\; .
\end{eqnarray}
The upper sign refers to an antisymmetric mode $x=-y$ whereas the
lower sign refers to a symmetric mode $x=y$. Stationary bifurcations
occur at $\sigma=0$. The symmetric mode then coincides with the
saddle-node bifurcation (\ref{SN_wt}) whereas the antisymmetric mode
terminates at the pitchfork bifurcation (\ref{AI_PFb}).\\

\noindent
Non-stationary Hopf bifurcations are possible if $\sigma=i\omega$. We
then have
\begin{eqnarray}
\label{lin_ai3}
\omega=\mp \beta \sin{\omega \tau}
\end{eqnarray}
and
\begin{eqnarray}
\label{lin_ai4}
{\bar X}_h= \frac{\beta}{2g}(\pm\cos{\omega \tau}-\gamma_1)\; .
\end{eqnarray}
One has physical solutions with a single-valued positive $\omega$ only
for the symmetric case (the lower signs) which reproduces our results
(\ref{HomHopfa}) and (\ref{HomHopfb}) for the symmetry preserving Hopf
bifurcation. For $\omega \tau\to 0$ the Hopf bifurcation moves towards
the saddle-node (\ref{SN_wt}) and coalesces with it at $\beta \tau=1$
in a Bogdanov-Takens point as described in Section~\ref{Sec-BT}.  For
$\omega \tau\to\pi$ the limiting value of ${\bar X}_{h}$ is ${\bar
X}_h=\beta(1-\gamma_1)/(2g)$ which coincides with the pitchfork
bifurcation $X_{PF}$ in another codimension-2 bifurcation. At this
bifurcation the Hopf bifurcation has a period $T = 2 \tau$ which
corresponds exactly to the inhomogeneous pitchfork bifurcation with
$p=\pi$ whereby every second pulse dies.

For values $\omega \tau \in [0,\pi)$ the Hopf bifurcation always comes
after the pitchfork bifurcation which has been numerically verified
with simulations of the full system (\ref{barkley}).\\

\begin{figure}
\label{Fig-Pitchsketch}
\centerline{
\psfig{file=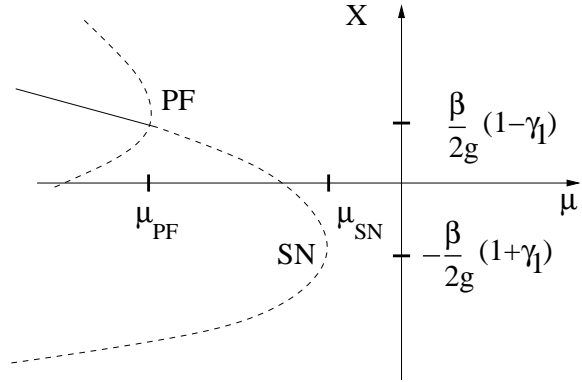,angle=0,width=3.0in}}
\caption{Sketch of the bifurcation diagram for a wave train
in a ring showing a stationary saddle-node bifurcation (SN) and a
subcritical pitchfork bifurcation (PF).}
\end{figure}

\noindent
This allows us to sketch the full bifurcation scenario for a wave
train in a periodic ring as depicted in Figure~\ref{Fig-Pitchsketch}.


\noindent
\section {Determination of the parameters of the normal form}
\label{Sec-Num1}
\vskip 5pt 

In this Section we determine the parameters of the normal form
(\ref{NF}) from numerical simulations of the full partial differential
equations (\ref{barkley}). We determine the free parameters $\nu$,
$\mu$, $g$, $\gamma$, $\gamma_1$ and $\beta_0$. We are then in the
position to test how well the normal form (\ref{NF}) reproduces the
solution behaviour of the full partial differential equation
(\ref{barkley}). We use here as equation parameters for
(\ref{barkley}) $a=0.22$, $u_s=0.1$ and $D=1$.\\

\noindent
The parameter $\nu$ which modifies the delay time $\tau$ due to the
finite width of a pulse is easily determined as a typical width of a
pulse in the parameter region of interest. We find $\nu=32$. We note
that there is some ambiguity in the determination of $\nu$, and one
may as easily justify $\nu \in [29,34]$.\\

\begin{figure}
\label{Fig-dtXeps}
\centerline{
\psfig{file=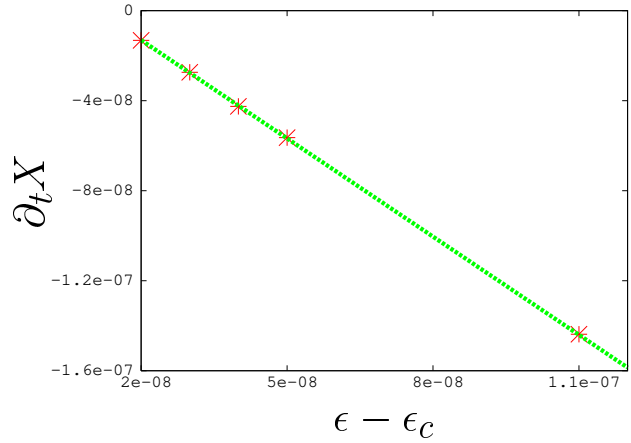,angle=-270,width=4.0in}}
\caption{The temporal derivative $\partial_t X$ at
the saddle-node versus $\epsilon-\epsilon_c$. The slope determines the
parameter $\alpha=\mu/(\epsilon-\epsilon_c)$. The stars are obtained
by numerically integrating the partial differential equation
(\ref{barkley}). The line is a least-square fit.}
\end{figure}

\noindent
The parameters $\mu$ and $g$ can be determined by studying the
isolated pulse with $L\to \infty$. The normal form reduces to
\begin{eqnarray}
\label{para1}
\partial_t X = -\mu - g X^2 \; .
\end{eqnarray}
We have $\mu=\alpha(\epsilon-\epsilon_c)$ with $\epsilon_c$ being the
critical $\epsilon$ at the saddle-node. Solutions of (\ref{para1}) are
obtained by quadrature
\begin{eqnarray}
\label{para1b}
X(t) &=& \sqrt{\frac{-\mu}{g}}\tanh(\sqrt{-\mu g}(t-t_0)) \quad
{\rm{for}} \quad  \mu<0 \nonumber\\
\quad X(t) &=&
-\sqrt{\frac{\mu}{g}}\tan(\sqrt{\mu g}(t-t_0))\quad \, \, \, \quad {\rm{for}} \quad
\mu>0 \; .
\end{eqnarray}
For small deviations from the saddle-node $X=0$ this solution may be
expanded to obtain $X(t)\approx \mu t$ which obviously corresponds to
the solution of equation (\ref{para1}) linearized around the
saddle-node. The solution (\ref{para1b}) has its inflection point at
the saddle-node $X=0$ where its slope is $\mu$. We can therefore
determine $\mu$ by measuring $\partial_t X$ at the inflection point
for different values of $(\epsilon-\epsilon_c)$. This allows us to
determine $\alpha$ via $\alpha=\mu/(\epsilon-\epsilon_c)$. In
Figure~\ref{Fig-dtXeps} we show the results of $\partial_t X$ versus
$(\epsilon-\epsilon_c)$. The numerical results are obtained by letting
a stable pulse which was created at some $\epsilon<\epsilon_c$, decay
in an environment with $\epsilon>\epsilon_c$. The relaxation then
allows us to determine the slope at the saddle-node. Using a
least-square fit we obtain $\alpha=1.455$.\\

\begin{figure}
\label{Fig-epsX}
\centerline{
\psfig{file=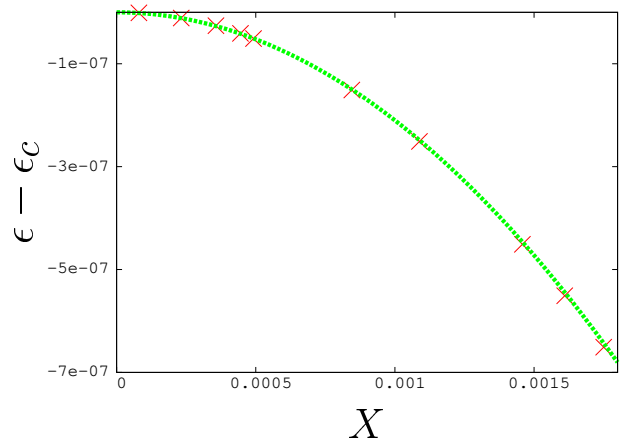,angle=-270,width=4.0in}}
\caption{Bifurcation parameter $\epsilon-\epsilon_c$ versus
amplitude $X$ close to the saddle-node of the isolated pulse. The
crosses are obtained by numerically integrating the partial
differential equation (\ref{barkley}). The line is a quadratic fit.}
\end{figure}

\noindent
The parameter $g$ can now be determined by looking at the stationary
problem $\partial_t X=0$. The behaviour of the amplitude of the
activator close to the saddle-node versus $\epsilon$ is depicted in
Figure~\ref{Fig-epsX}. It clearly demonstrates quadratic behaviour
typical for saddle-nodes. The normal form for the saddle-node of an
isolated pulse (\ref{SN0}) yields $(\epsilon - \epsilon_c) =
(g/\alpha) X^2$ which we can use upon using the above measured value
of $\alpha$ to obtain $g=0.31$ from a least square fit.\\

\noindent
To determine the missing parameters $\beta_0$, $\gamma$ and $\gamma_1$
we need to study a pulse in a periodic ring of finite length $L$. The
self-interaction of the pulse with its own inhibitor modifies the
saddle-node bifurcation as discussed in Section~\ref{Sec-NF}. We will
look here at the symmetry preserving Hopf bifurcation. To study the
Hopf bifurcation we study the circulation of a single pulse in a
periodic ring instead of a wave train consisting of more than one
pulse. In the latter case the subcritical pitchfork bifurcation sets
in before the symmetry preserving Hopf bifurcation.

\begin{figure}
\label{Fig-XHHom}
\centerline{
\psfig{file=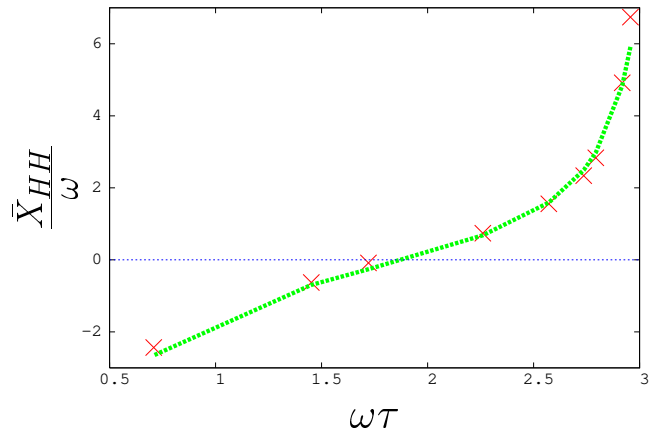,angle=-270,width=4.0in}}
\caption{Plot of ${\bar{X}}_{HH}/\omega$ versus $\omega
\tau$. The crosses depict results from numerically integrating the
full partial differential equation (\ref{barkley}). The line
represents the analytical expression (\ref{XhOm}) of our normal
form. This fit is achieved by choosing $\gamma_1=0.31$.}
\end{figure}

Combining the expressions for the angular frequency $\omega$ and the
deviation ${\bar{X}}_{HH}$ of the pulse amplitude from the
saddle-node at the bifurcation point, (\ref{HomHopfa}) and
(\ref{HomHopfb}), we can eliminate the so far undetermined parameter
$\beta$ to determine $\gamma_1$. We obtain
\begin{eqnarray}
\label{XhOm}
\frac{{\bar{X}}_{HH}}{\omega}=-\frac{1}{2g}(\cot(\omega \tau) +
\gamma_1\sin^{-1}(\omega \tau)) \; .
\end{eqnarray}
In Figure~\ref{Fig-XHHom} we show a plot of numerically obtained
values for the quantity ${\bar{X}}_{HH}/\omega$ by integrating the
partial differential equation (\ref{barkley}), and the result of our
normal form (\ref{XhOm}). In the numerical simulations of
(\ref{barkley}) we measured the frequency of the subcritical Hopf
bifurcation $\omega$ and ${\bar{X}}_{HH}$ which is the difference
between the amplitude at the symmetry preserving Hopf bifurcation and
the saddle-node value of the isolated pulse. Expression (\ref{XhOm})
matches the numerical simulations well for $\gamma_1=0.31$.\\

\noindent
We can now determine the parameters $\beta_0$ and $\gamma$ by looking
at the shift of the bifurcation parameter $\mu$ and the shift of the
critical amplitude $X$ at the saddle-node for finite wavelength $L$
with respect to the values at the saddle-node for an isolated pulse
with infinite wavelength. The saddle-node is shifted with respect to
the isolated pulse with $\mu_{SN}=X_{SN}=0$. We express the shifted
finite-$L$ saddle-node by
\begin{eqnarray*}
\mu+\delta=-g(X-\xi)^2 \; ,
\end{eqnarray*}
where $\delta=\delta(L)$ expresses the shift in the bifurcation
parameter at the saddle-node and $\xi=\xi(L)$ expresses the shifted
value of the amplitude when compared to the isolated pulse. 

\begin{figure}
\label{Fig-shiftpara}
\centerline{
\psfig{file=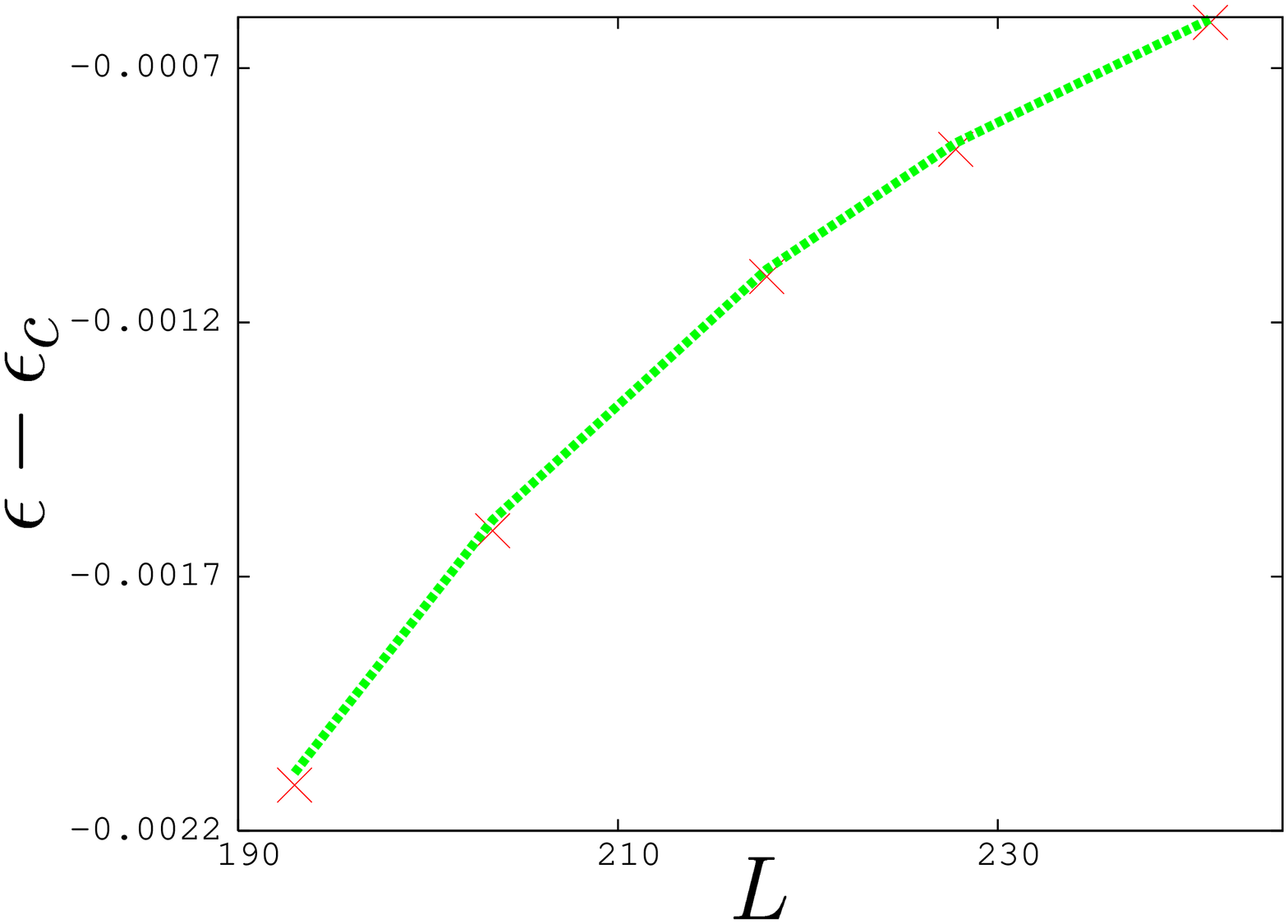,angle=-270,width=4.0in}}
\caption{Shift of the bifurcation parameter
$(\epsilon-\epsilon_c)$ at the saddle-node as a function of
$L$. Crosses depict results from a numerical simulation of the full
partial differential equation (\ref{barkley}). The line represent the
analytical expression (\ref{shifta}) of our normal form. Agreement of
the two curves is achieved for $\gamma\beta_0=0.115$.}
\end{figure}

\begin{figure}
\label{Fig-shiftX}
\centerline{
\psfig{file=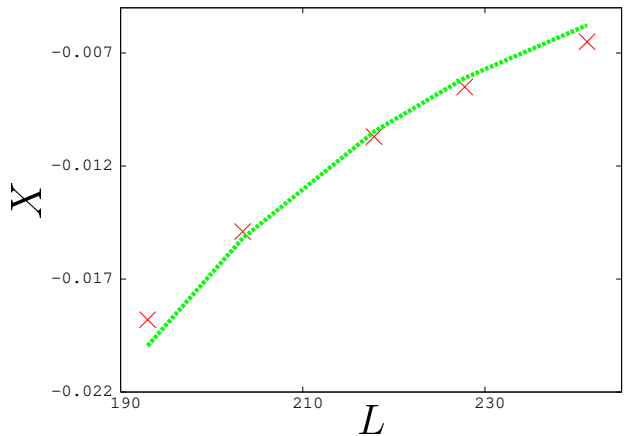,angle=-270,width=4.0in}}
\caption{Shift of the amplitude deviation $X$ at the
saddle-node as a function of $L$. Crosses depict results from a
numerical simulation of the full partial differential equation
(\ref{barkley}). The line represent the analytical
expression (\ref{shiftb}) of our normal form. Agreement of the two
curves is achieved for $(\beta_0/2g)(1+\gamma_1) = 0.91$.}
\end{figure}

The values for $\delta(L)$ and $\xi(L)$ can be measured by numerically
solving (\ref{barkley}) and determing the finite wavelength induced
saddle-node. We did so by expressing (\ref{barkley}) in travelling
wave coordinates and treating the problem as a boundary value
problem. The thereby obtained values for $\delta(L)$ and $\xi(L)$ have
to be compared with the corresponding expressions of the normal form.
In the normal form (\ref{NF}) the finite wavelength induced shift is
represented by the finite length correction
\begin{eqnarray*}
\label{V}
V(t-\tau)=\beta_0\exp(-a\epsilon(L-\nu)/c)\left(\gamma +
X(t-\tau)+\gamma_1X(t) \right) \; .
\end{eqnarray*}
Neglecting ${\cal{O}}(\xi^2)$ which is justified for not too large
deviations from the isolated pulse, we obtain
\begin{eqnarray}
\label{shifta}
\delta &=& \gamma \beta_0 \; \exp(-\epsilon a (L-\nu)/c)\\
\label{shiftb}
\xi   &=& \frac{\beta_0}{2g}(1+\gamma_1)\; \exp(-\epsilon a (L-\nu)/c)\; ,
\end{eqnarray}
where $\nu$ and $\gamma_1$ had already been determined. Note that the
${\cal{O}}(\xi^2)$-term, $g\xi^2=\beta^2(1+\gamma_1)/4g$ appears, of
course, correctly in the shift of the bifurcation parameter $\mu_{SN}$
in (\ref{SN_wt}). Although the inclusion of $\gamma_1\neq 0$ is not
necessary for the existence of a Hopf bifurcation (see
(\ref{HomHopfa},\ref{HomHopfb})), it is significant to obtain good
quantitative agreement. Whereas for the stationary bifurcations the
inclusion of $\gamma_1$ is in effect a redefinition of $\beta$, it is
vital in the case of the Hopf bifurcation because it allows for a
decoupled dependence of the frequency $\omega$ and the amplitude
${\bar X}_{HH}$ on $L$.
In Figure~\ref{Fig-shiftpara} we show a plot of the bifurcation
parameter shift as a function of length $L$. Agreement of numerically
obtained values from a simulation of the full partial differential
equation (\ref{barkley}) with the expression derived from our normal
form (\ref{shifta}), which implies
$\delta=\alpha(\epsilon-\epsilon_c)$ is assured provided
$\gamma\beta_0=0.115$. We recall that $\epsilon_c$ is the critical
refractoriness at the saddle-node of an isolated pulse. In
Figure~\ref{Fig-shiftX} we show a plot of the pulse amplitude shift as
a function of length $L$. Agreement of numerically obtained values
from a simulation of the full partial differential equation
(\ref{barkley}) with the expression derived from our normal form
(\ref{shiftb}) is given provided $(\beta_0/2g)(1+\gamma_1) = 1.1$.
Combining these results we can solve for $\beta_0$ and $\gamma$ and
obtain $\gamma = 0.32$ and $\beta_0=0.53$.
This finalizes the determination of the free parameters of the normal
form (\ref{NF}).\\

\begin{figure}
\label{Fig-XHH}
\centerline{
\psfig{file=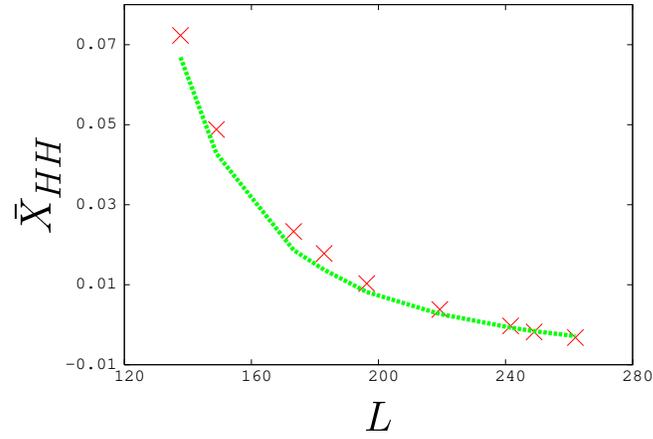,angle=-270,width=4.0in}}
\caption{Amplitude deviation at the Hopf bifurcation ${\bar{X}}_{HH}$ as a
function of $L$. Crosses depict results from a numerical simulation of
the full partial differential equation (\ref{barkley}). The line
represent the analytical expression (\ref{HomHopfb}) of our normal
form.}
\end{figure}

\begin{figure}
\label{Fig-om}
\centerline{
\psfig{file=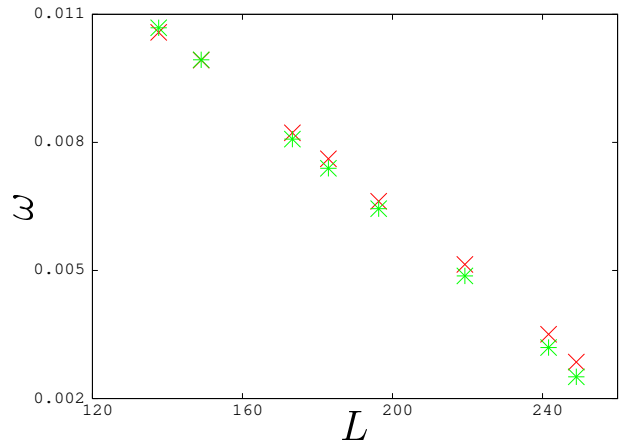,angle=-270,width=4.0in}}
\caption{Angular frequency at the Hopf bifurcation $\omega$ as a
function of $L$. Crosses depict results from a numerical simulation of
the full partial differential equation (\ref{barkley}). Stars were
obtained by solving the analytical expression (\ref{HomHopfa}) of our
normal form for $\omega$.}
\end{figure}

\noindent
We are now in the position to check the validity of our normal form by
testing whether the normal form (\ref{NF}) with the above determined
parameters is able to reproduce observations of the numerical
simulation of the full partial differential equation
(\ref{barkley}). We already noted in Section~\ref{Sec-BT} qualitative
agreement i.e. the correct bifurcation behaviour. Here we show
quantitative agreement with the behaviour of (\ref{barkley}).\\ In
particular we look at the symmetry preserving Hopf bifurcation
described in Section~\ref{Sec-BT}. In Figure~\ref{Fig-XHH} and
Figure~\ref{Fig-om} we show a comparison of the analytical results for
the frequency and the amplitude at the Hopf bifurcation,
(\ref{HomHopfa}) and (\ref{HomHopfb}), with results obtained from
numerically integrating (\ref{barkley}). The Figures show good
agreement. Note that the parameters $\beta_0$ and $\nu$ were not
determined by fitting data representing the symmetry preserving Hopf
bifurcation and henceforth the two figures, Figure~\ref{Fig-XHH} and
Figure~\ref{Fig-om}, are indeed predictions.\\

\begin{figure}
\label{Fig-XPF}
\centerline{
\psfig{file=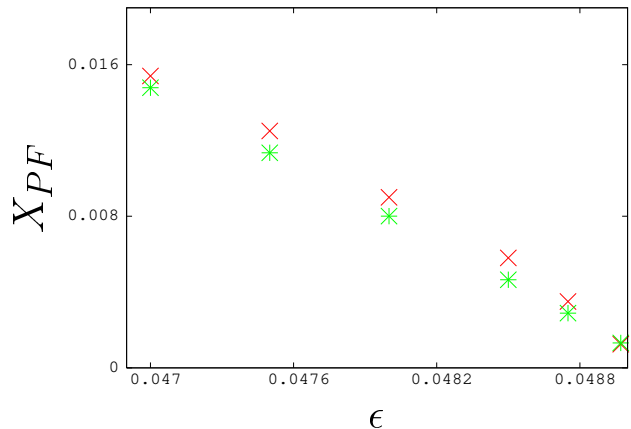,angle=-270,width=4.0in}}
\caption{Amplitude deviations at the pitchfork
bifurcation ${\bar{X}}_{PF}$ as a function of $\epsilon$. Crosses
depict numerical results from integrating the full partial
differential equation (\ref{barkley}). Stars show the corresponding
values calculated by using the normal form result (\ref{AI_PFb}).}
\end{figure}

\noindent
Figure~\ref{Fig-XPF} shows a comparison of numerical simulations of
(\ref{barkley}) and our analytical results (\ref{AI_PFb}) for the
spatially inhomogeneous pitchfork bifurcation. For the determination
of the parameters of the normal form we have not used any fitting
which involved results from the spatially inhomogeneous pitchfork
bifurcation. The agreement in Figure~\ref{Fig-XPF} therefore
demonstrates that the normal form can indeed be used to obtain
quantitative agreement and make quantitative predictions. 

\medskip
\section{Summary and discussion}
\label{Sec-Disc}
We have constructed a normal form for travelling waves in
one-dimensional excitable media which takes the form of a delay
differential equation. The construction is based on the well-known
observation that the interaction of a pulse with the inhibitor of the
preceding pulse modifies the generic saddle-node bifurcation of an
isolated pulse. The normal form (\ref{NF}) exhibits a rich bifurcation
behaviour which we could verify by numerically simulating the partial
differential equation (\ref{barkley}). Besides the well known
saddle-node bifurcations for isolated pulses and for periodic wave
trains the normal form also exhibits a symmetry preserving Hopf
bifurcation and a symmetry breaking, spatially inhomogeneous pitchfork
bifurcation. Moreover, the normal form shows that the saddle-node and
the Hopf bifurcation are an unfolding of a Bogdanov-Takens point as
previously suggested in \cite{Knees92,GottwaldKramer04}. The symmetry
preserving Hopf bifurcation is found to occur before the saddle-node
bifurcation for a single pulse in a ring. For a wave train consisting
of several pulses in a ring, the Hopf- and the saddle-node
bifurcations occur after the symmetry breaking pitchfork bifurcation
in which every second pulse dies. We could verify these scenarios in
numerical simulations of the modified Barkley-model (\ref{barkley})
and the Fitzhugh-Nagumo equations \cite{FHN}. These bifurcations have
been observed before but had previously not been described within a
unified framework of one normal form.\\ We were able to determine the
parameters of the normal form from numerical simulations of the
partial differential equation (\ref{barkley}). Using these numerically
determined parameters we showed excellent agreement between the normal
form and the full partial differential equation (\ref{barkley}). We
could quantitatively describe the symmetry preserving Hopf bifurcation
and the inhomogeneous pitchfork bifurcation. Moreover, we were able to
quantify the Bogdanov-Takens bifurcation.\\

\noindent
The symmetry preserving Hopf bifurcation has been studied intensively
before. It was observed numerically for example for the Barkley-model
\cite{Knees92}. Interest has risen in the Hopf bifurcation in the
context of cardiac dynamics because it leads to propagation failure of
a single pulse on a ring. It is believed to be related to a phenomenon
in cardiac excitable media which goes under the name of {\it
alternans}. Alternans describe the scenario whereby action potential
durations are alternating periodically between short and long
periods. The interest in alternans has risen as they are believed to
trigger spiral wave breakup in cardiac tissue and ventricular
fibrillation
\cite{Nolasco,Courtemanche,Karma93,Karma_A,Fenton02}. Besides
numerical investigations of the Barkley model \cite{Knees92}, the
modified Barkley model \cite{GottwaldKramer04}, the Beeler-Reuter
model \cite{QuanRudy,Courtemanche,Karma94,Courtemanche96}, the
Noble-model \cite{Karma93,Karma94} and the Karma-model \cite{Karma93},
where a Hopf bifurcation has been reported, there have been many
theoretical attempts to quantify this bifurcation for a single-pulse
on a ring. Since the pioneering work \cite{Nolasco} alternans have
been related to a period-doubling bifurcation. It was proposed that
the bifurcation can be described by a one-dimensional return map
relating the action potential duration ($APD$) to the previous
recovery time, or diastolic interval ($DI$), which is the time between
the end of a pulse to the next excitation. A period-doubling
bifurcation was found if the slope of the so called restitution curve
which relates the $APD$ to the $DI$, exceeds one. A critical account
on the predictive nature of the restitution curve for period-doubling
bifurcations is given in \cite{Fenton99,Fox02}. In \cite{Karma94} the
instability was analyzed by reducing the partial differential equation
describing the excitable media to a discrete map via a reduction to a
free-boundary problem. In \cite{GottwaldKramer04} the Hopf bifurcation
could be described by means of a reduced set of ordinary-differential
equations using a collective coordinate approach. In
\cite{Courtemanche,Courtemanche96} the bifurcation was linked to an
instability of a single integro-delay equation. The condition for
instability given by this approach states - as in some previous
studies involving one-dimensional return maps - that the slope of the
restitution curve needs to be greater than one. However, as evidenced
in experiments \cite{Hall} and in theoretical studies
\cite{Fenton99,Fox02} alternans do not necessarily occur when the
slope of the restitution curve is greater than one. In further studies
it would be interesting to see how our criterion $\beta\tau\ge 1$ is
related to that condition. Note that $\nu$ is related to the $APD$ and
$\tau=(L-\nu)/c$ to the recovery time $DI$.\\
\noindent
In the context of alternans the Hopf bifurcation has been described as
a supercritical bifurcation
\cite{Courtemanche,Karma93,Karma94,Courtemanche96} (although their
occurrence is related to wave break up \cite{Karma94}). Further study
will explore whether the subcritical character of the Hopf bifurcation
we find is model-dependant or indeed generic.\\

\noindent
To go beyond the case of a single pulse circulating in a ring,
periodically stimulated excitable media have been studied in the
context of alternans
\cite{Guevara81,Lewis90,Echebarria01,Hastings00,Fox02,Henry05}. In
\cite{Guevara81,Fox02} one-dimensional maps were developed to
study the Hopf bifurcation and the transition to conduction blocks. In
\cite{Echebarria01} a nonlocal partial differential equation has been
proposed to study spatiotemporal dynamics of alternans. It would be
interesting for further studies to see how the transition to
conduction blocks explored in these paced cardiac excitable media
\cite{Hastings00,Echebarria01,Fox02,Henry05} 
can be described by the spatially inhomogeneous pitchfork bifurcation
we found in our normal form. The pitchfork bifurcation however does
not require a fixed pacing site and does not require external pacing
but rather is dynamically induced. This may aid in investigating the
formation of conduction blocks purely as a dynamical phenomenon of
wave trains.\\
\noindent 
An interesting scenario in our normal is the coalescence of the
spatially inhomogeneous pitchfork bifurcation with the Hopf
bifurcation when $\omega \tau = \pi$. This condition implies
$T=2\tau$. Then the Hopf frequency is in resonance with the spatial
instability in which every second pulse dies. Connections to alternans
of this scenario are planned for further research.\\

\noindent
Ideally, one would like to deduce the normal form directly from a
model for excitable media and determine its parameters without relying
on numerical simulations of a particular excitable medium. One initial
path along that avenue could be to use the non-perturbative approach
developed in \cite{GottwaldKramer04} to determine the parameters. This
method was developed to study critical wave propagation of single
pulses and pulse trains in excitable media in one and two
dimensions. It was based on the observation that close to the
bifurcation point the pulse shape is approximately a bell-shaped
function. Numerical simulations show that this is the case for the
Barkley model (\ref{barkley}) close to the saddle-node bifurcation. A
test function approximation that optimises the two free parameters of
a bell-shaped function, i.e. its amplitude and its width, allows us to
find the actual bifurcation point, $\epsilon_c$, and determine the
pulse shape for close-to-critical pulses at excitabilities near
$\epsilon_c$. This method has so far also been successfully applied to
other non-excitable reaction-diffusion equations
\cite{Menon05,Cox05}. To apply the method for our purpose is planned
for future work.\\


\medskip

{\underbar{\bf Acknowledgements }} G.A.G gratefully acknowledges
support by the Australian Research Council, DP0452147. G.A.G. would
like to thank Bj\"orn Sandstede for kindly helping with producing some
of the numerical saddle-node values using AUTO. Initial parts of the
simulations were done with the XDim Interactive Simulation Package
developed by P. Coullet and M. Monticelli.
\vskip 5pt
\vfill\eject


\newpage


\end{document}